# MINIJETS AT SMALL $x$

P.V. Landshoff, DAMTP, University of Cambridge

**Abstract**

Nonperturbative pomeron exchange at high energy includes minijet production. At moderate $Q^2$ it is responsible for the small-$x$ behaviour of $\nu W_2$. Hence minijet production should be a feature of deep inelastic scattering at small $x$.

Minijets are jets whose transverse momentum is so small that they are difficult, or even impossible, to detect experimentally. Even so, theorists may consider them, and their production is well worth measuring, as it can give important information about the way in which perturbative and nonperturbative effects combine in QCD. Figure 1 shows data for the photoproduction total cross-section. The curves are extrapolations of two different fits to the low-energy data. The one that goes through the two HERA points is a Regge fit[1], which consists of two terms. One behaves as $s^{0.45}$ and corresponds to $\rho, \omega, f_2, a_2$ exchange, and the other corresponds to soft pomeron exchange and behaves as $s^{0.08}$. (Both are effective powers, but they vary only <u>very</u> slowly with $s$). The upper curve[2] is obtained by integrating the inclusive cross-section $d\sigma/dp_T$ for minijet production down to $p_T^{MIN} = 1.4 \,\text{GeV}/c$; it is interesting to consider why it is so far above the HERA data[2)3)].

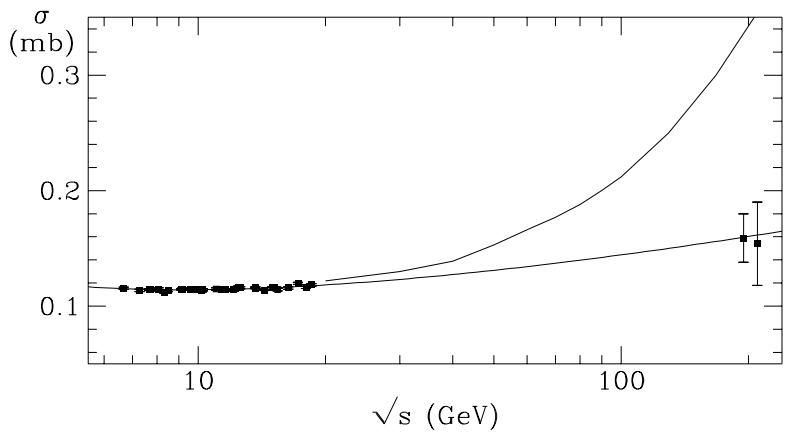
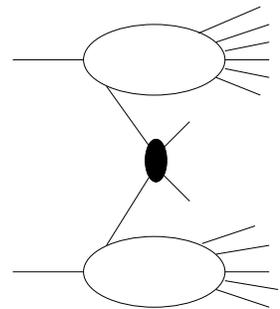

Figure 1: $\gamma p$ total cross section       Figure 2: Hard-scattering mechanism

In lowest-order QCD, $d\sigma/dp_T$ is calculated in the usual way from figure 2, which involves a hard scattering in the middle of the diagram and structure functions for the photon and the proton. Because the typical scale where nonperturbative effects become important is about 1 GeV, I would expect figure 2 still to reproduce $d\sigma/dp_T$ reasonably accurately at $p_T = 1.4$ GeV/c, in the sense that any additional $K$-factor that may be necessary will probably be somewhere between $\frac{1}{2}$ and 2. If I integrate the inclusive cross-section for the production of a pair of minijets down to $P_T^{MIN}$, I obtain[4]

$$\int_{p_T^{MIN}} dp_T \frac{d\sigma^{\text{PAIR}}}{dp_T} = \bar{n} \rho \sigma^{\text{TOT}} \quad (1)$$

where $\rho$ is the fraction of events that have a pair of minijets, and $\bar{n}$ is the average number of minijets pairs in those events. From figure 1, for $p_T^{MIN} = 1.4$ we find that $\bar{n}\rho \approx 2$ at HERA energies. Since, by definition, $\rho \leq 1$, this means that $\bar{n} > 1$.

How, then, can one generate events that have more than one pair of minijets? One way is for more than one pair of partons to undergo a hard scattering. This is certainly very important in nucleus-nucleus scattering[5], but I suspect that it is much less so in $\gamma p$ or $pp$ collisions.

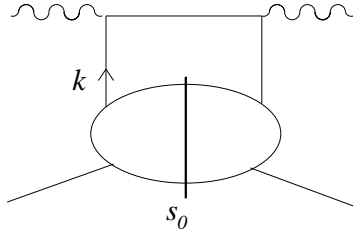

Figure 3: Parton model for $\nu W_2$

Another mechanism is closely related to the fact[6] that $\nu W_2$ is Regge behaved at small $x$. Consider the simple parton model, figure 3. From elementary kinematics, the squared invariant mass $s_0$ of the parton fragments left behind when the parton $k$ is pulled out is

$$s_0 \sim -\frac{k^2 + k_T^2}{x} \qquad (2)$$

and so it becomes large at small $x$. Since the lower proton/parton amplitude in figure 3 is a strong-interaction amplitude, it should be dominated by Regge exchanges when its energy variable $s_0$ becomes large, so that its behaviour is a sum of two terms $s^{\alpha(0)}$ with $\alpha(0) - 1 = 0.45$ and $0.08$ respectively. When this is inserted into the calculation of figure 3, the result is that $\nu W_2$ at small $x$ behaves as a sum of terms $x^{1-\alpha(0)}$. This fits well[7] to the data[8] from NMC, as is shown in figure 4. The fit here is

$$\nu W_2 = 0.32\, x^{-0.08} \left(\frac{Q^2}{Q^2 + a}\right)^{1.08} + 0.10\, x^{0.45} \left(\frac{Q^2}{Q^2 + b}\right)^{0.55} \qquad (3)$$

with $a = (750\,\text{MeV})^2$ and $b = (110\,\text{MeV})^2$. Multiplying this form by $4\pi^2\alpha/Q^2$ and setting $Q^2 = 0$, we retrieve also the lower curve in figure 1.

In figure 2, the $x$-values of the two partons that undergo the hard collision are of order $p_T^{MIN}/\sqrt{s}$, and so are small. Hence the invariant masses of the two systems of remnant hadrons, of the photon and the proton, are both large. So it is quite possible[4] that these two systems of hadrons themselves contain minijets. That is, from figure 2 we find that in minijet events the average number $\bar{n}$ of minijets pairs is greater than 1.

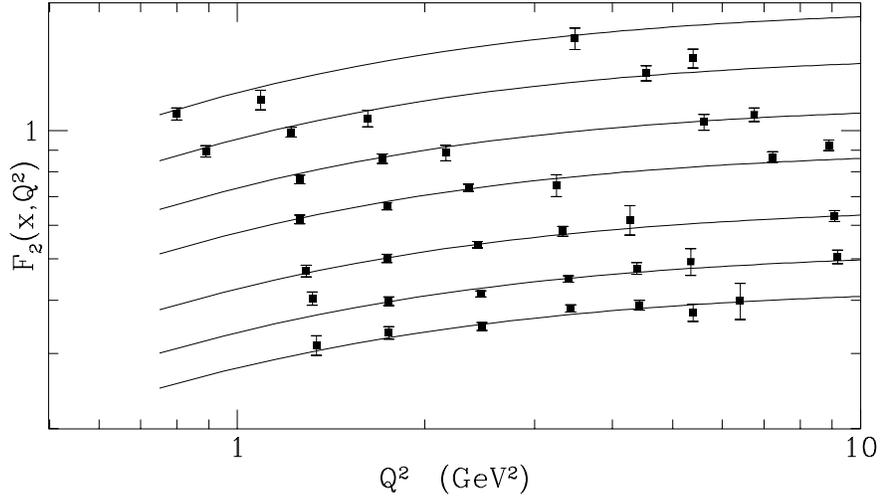

Figure 4: Simple fit to NMC data for $\nu W_2$ for values of $x$ between .008 and .07

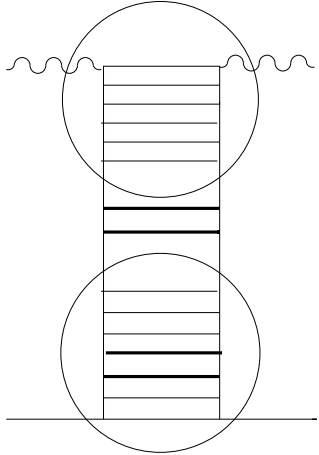

Figure 5: Ladder model for the square of figure 2; the circles denote the structure functions

This is a basically nonperturbative effect. Figure 5 shows the square of figure 2 in a ladder model. The two circles denote the two structure functions, and between them there is the hard scattering shown in figure 2, with moderately high $p_T$ ($p_T > p_T^{MIN}$) round the loop. However, one or more of the other loops within the structure functions, may well also have $p_T > p_T^{MIN}$. At small $x$ there is no transverse momentum ordering, so the "hot" loops are likely to be separated by loops that do not carry high $p_T$.

We can calculate this[9] in a simple factorising model. Because at small $x$ the structure functions are dominated by Regge exchange, with pomeron exchange the most important component, we may represent figure 5 as shown in figure 6a, which has two nonperturbative pomerons and a central perturbative hard scattering. At small $p_T$, the hard scattering $gg \to gg$ is the most important.

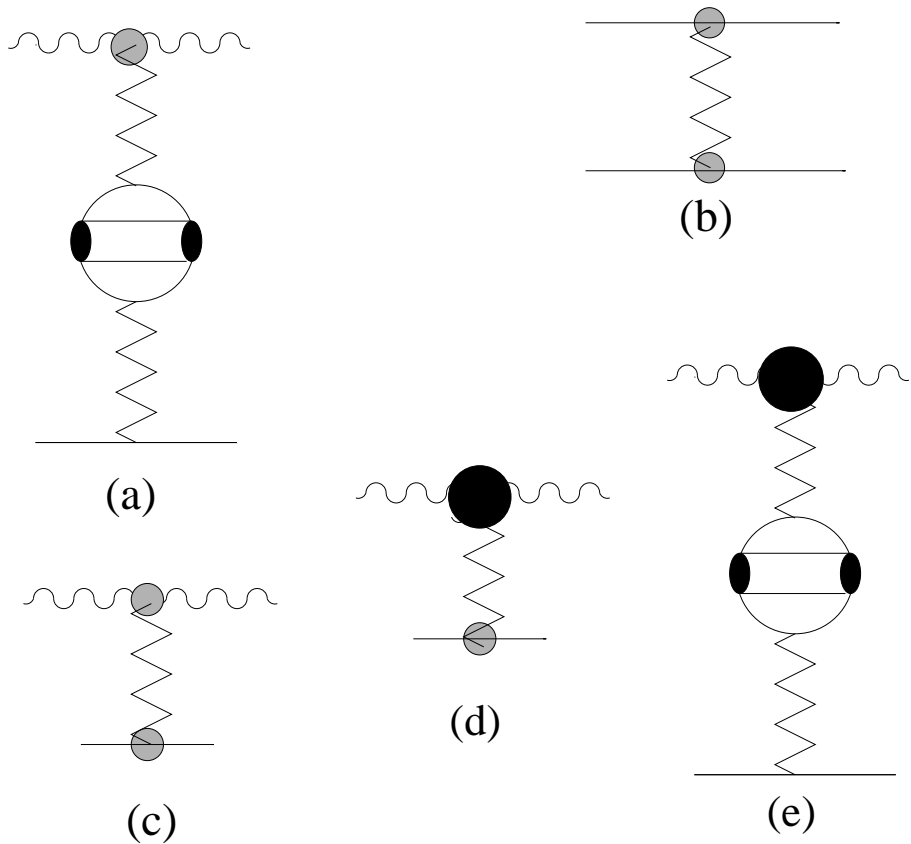

Figure 6: Simple factorisable model

As I have indicated, the two pomerons may themselves include additional hard scatterings within them. To calculate figure 6a, we need the couplings, $\beta_\gamma$, $\beta_p$ and $\beta_g$ of the pomeron to the photon, proton and gluon. $\beta_p$ is determined from $\sigma^{\text{TOT}}(pp)$, which corresponds to figure 6b and behaves as $\beta_p^2 s^{0.08}$ at large $s$. Then $\beta_\gamma$ is determined from $\sigma^{\text{TOT}}(\gamma p)$, which is calculated from figure 6c and behaves as $\beta_\gamma \beta_p s^{0.08}$. Comparing figure 6a with the upper curve in figure 1, we then determine $\beta_g$.

We may now use this to predict minijet production in deep inelastic lepton scattering at small $x$. The structure function $\nu W_2$ is determined from figure 6d and behaves as $C(Q^2)\beta_p x^{-0.08}$, where $C(Q^2)$ is the upper bubble, and then the minijet production is calculated from figure 6e. The result[9] is given in figure 7, which shows the fractions of events containing minijets with $p_T > 3\text{GeV}/c$ and $p_T > 5\text{GeV}/c$. The bands correspond to a nonperturbative scale associated with the pomeron-gluon coupling ranging between 0.7 and 1.4 GeV. At small and moderate values of $Q^2$ the fraction is predicted to be independent of $Q^2$. The minijets will be uniformly distributed in rapidity in the central region. I expect that the results in figure 7 are actually lower limits. This is because they assume that, at small $x$, $\nu W_2$ is dominated by soft pomeron exchange. However, if one sets $Q^2 = 8.5$ GeV$^2$ and extrapolates the fit (3) down to $x = 0.0002$, it falls significantly below the measurements reported by HI. The precise reason for this is not yet agreed, but whatever it is one might expect it to yield additional minijet production.

To conclude, I have explained that minijet production is part of pomeron exchange. Further experimental data will help us better to understand pomeron dynamics.

*This research is supported in part by the EC Programme "Human Capital and Mobility" Network "Physics at High Energy Colliders" contract CHRX-CT93-0537 (DG 12 COMA)*

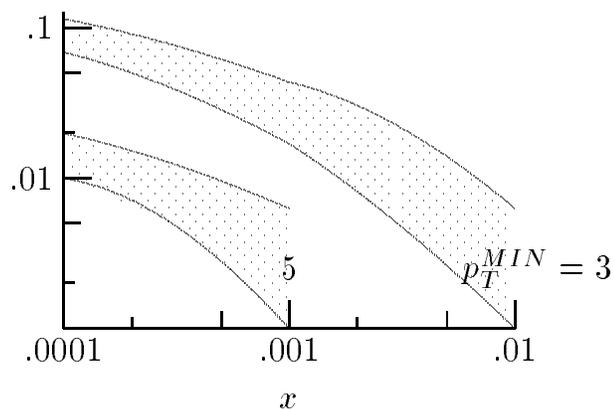

Figure 7: Fraction of deep inelastic scattering events that contain minijets